\documentclass[11pt]{article}
\usepackage[margin=1in]{geometry}

\usepackage{amsmath}
\usepackage{amsthm}
\usepackage{amssymb}
\usepackage{mathtools}
\usepackage{derivative}
\usepackage{algpseudocode}
\usepackage{algorithm}

\usepackage{graphicx}
\usepackage{booktabs}
\usepackage{tabularx}

\usepackage{xcolor}     %
\usepackage{steinmetz}
\usepackage{lipsum}
\usepackage{tikz}

\usepackage{natbib}
\usepackage{hyperref}

\usepackage{subfig}
\hypersetup{
  colorlinks=true,
  linkcolor=blue,
  citecolor=blue,
  urlcolor=blue
}

\usepackage{setspace}
\DeclareMathOperator*{\argmax}{argmax}
\DeclareMathOperator*{\argmin}{argmin}

\newtheorem{assumption}{Assumption}
\newtheorem{theorem}{Theorem}

\title{Proximal Individualized Functional Treatment Regimes}

\author{
Zhuoxin Long\textsuperscript{1},
Xiaoke Zhang\textsuperscript{1}\thanks{Corresponding author: xkzhang@gwu.edu}
\\[1ex]
\small \textsuperscript{1}Department of Statistics,
George Washington University, Washington, DC, USA
\\
}

\date{}

\begin{document}

\maketitle

\begin{abstract}
{Estimating individualized treatment regimes (ITRs) is fundamental in data-driven personalized decision-making problems, such as precision medicine. Most of the ITR literature either focuses on categorical/continuous treatments or assumes no unmeasured confounding. In this paper, we make the first attempt to estimate the optimal individualized functional treatment regime (IFTR) for observational data where the treatment is a function and unmeasured confounding is present. We establish an identification result for a class of IFTRs under the proximal causal inference framework. Based on the identification result, we develop an algorithm of finding the optimal IFTR. The appealing practical performance of the proposed method is demonstrated by a simulation study. The proposed method is applied to an accelerometry dataset collected by the US National Health and Nutrition Examination Survey to find the optimal physical activity distribution for the best of the Triglyceride-Glucose index. }
\end{abstract}


\section{Introduction}

\noindent %
Identifying personalized treatment regimes for patients based on their individual characteristics in observational studies has gained great interest in recent decades. 
Typically, one wants to identify an optimal treatment regime that maximizes the mean health outcome among patients under its implementation. A vast number of statistical methods have been proposed for this objective
\citep[e.g.,][]{brinkley2010generalized,qian2011performance,zhao2012estimating, zhang2012estimating, zhang2012robust, laber2015tree,shi2018high,chernozhukov2019semi}. 
The majority of these works focus on binary or multi-categorical treatments, while fewer works study 
continuous treatments \citep[e.g.,][]{chen2016personalized, kallus2018policy,chernozhukov2019semi, zou2022counterfactual, schweisthal2023reliable}.  Recently, as data collection techniques have become more advanced, the functional treatment, where the treatment is a function/curve, has attracted increasing attention in observational causal inference \citep[e.g., ][]{ zhang2021covariate, zhao2025causal, tan2025causal,long2025learning, wang2026flexible, jiang2026estimating}. 
However, to date, there is limited work 
that estimates the optimal individualized treatment regime for functional treatments.  
To the best of our knowledge, \citet{Lin2026RL}, developed under an offline reinforcement learning framework, is the only closely related work. In this paper, we aim to narrow this gap and develop a new methodology of optimizing the individualized functional treatment regime (IFTR) in the presence of unmeasured confounding.

Our interest in this problem is motivated by accelerometry data collected by the National Health and Nutrition Examination Survey (NHANES) during 2013-2014. The data contains the 24-hour physical activity (PA) of participants measured during seven consecutive days by accelerometers.
The full trajectory of the PA data not only provides an objective measure of the PA, but also contains substantial and valuable information for free-living PA, including the volume, frequency, distribution, types of PA, etc. 
In this paper, we use the PA distributions derived from the accelerometry data as functional treatments, and we are interested in recommending individualized PA distributions for each participant for the best of his/her Triglyceride-Glucose index.

A great number of studies have utilized the accelerometry data to investigated the relationship between PA and health outcomes. 
Most PA studies reduce accelerometry data to one or more summary measures, such as total PA, light PA, moderate-to-vigorous PA, sedentary time, etc \citep{schmid2015associations,fishman2016association}.
Although this approach is simple to implement, it results in information loss and relies on the identification of cutoff points to define different types of PA. %
Lately, there have been attempts to treat the entire PA trajectory or PA distribution as a functional variable and investigate its correlational relationship with health outcomes using functional data analysis methods \citep[e.g.,][]{matabuena2023distributional, rogovchenko2024scalar, matabuena2025predicting}. 
More recently, 
\citet{long2025learning}, 
\citet{wang2026flexible}, \citet{jiang2026estimating}, and \citet{barnard2026causal} 
utilize the PA trajectory or PA distribution as the functional treatment and 
estimate its causal effect on a scalar outcome.
Despite the increasing interest in treating the PA distribution as a functional treatment, there has been essentially
no work on recommending the optimal PA distribution tailored to individual characteristics, 
which can be formulated as an IFTR learning problem.
The only exception, to the best of our knowledge, is \citet{Lin2026RL} 
that 
learns optimal 90-day step-count distributions associated with cardiometabolic risk.

One issue regarding the causal relationship between PA and the Triglyceride-Glucose index is that the relationship may be confounded by various factors including demographic, diet, lifestyle, socioeconomic, genetic factors, etc. While most of these factors have been measured in NHANES 2013-2014, the information of other potential confounders such as genetic factors are not collected. Thus, the typical unconfoundedness assumption is likely to be violated in this data and ITRs derived under this assumption can be therefore misleading. Although \citet{Lin2026RL} provides an important step toward learning individualized regimes for functional treatments, its offline reinforcement learning framework relies on a Markov decision process assumption with no unmeasured confounding. Thus, existing methods remain insufficient for learning ITRs from observational accelerometry data when unmeasured confounding may be present.

To address unmeasured confounding in ITR studies, 
a popular method 
is to use the proximal causal inference framework recently formulated by \citet{miao2018identifying} and \citet{tchetgen2024introduction}. 
Assuming that a rich set of proxies have been collected, and can be correctly allocated into three types: outcome-inducing proxies, treatment-inducing proxies and common causes of treatment and outcome, the framework can identify the mean treatment effect via a bridge function under the existence of unmeasured variables. This framework was 
adopted by \citet{qi2024proximal} to identify and 
learn optimal ITRs for a binary treatment.
\citet{shen2023optimal} extended this work by proposing a new class of ITRs that map all collected proxies to the treatment space, 
thereby further generalizing the regimes considered in \citet{qi2024proximal}.
In addition to binary treatments, there are some applications of this framework to ITR estimations for continuous treatments \citep[e.g.,][]{chen2023unified,li2026reinforcement}. However, these works focus on the regime of multiple time points 
and is thus out of the scope of this paper. 
To date, existing applications to ITRs only focus on categorical and continuous treatments, and there are no ITR methods for functional treatments in the presence of unmeasured confounding.
It is worth mentioning that the instrumental variable has also been employed to 
address unmeasured confounding in ITR learning \citep[e.g.,][]{cui2021semiparametric,qiu2021optimal}
However, 
the applicability of this approach is limited
given the restrictions on the 
binary treatment and instrumental variables \citep{qi2024proximal}.

In this paper, we adopt the proximal causal inference framework to account for unmeasured confounding in IFTR learning. We extend the value identification result in \citet{qi2024proximal} to functional treatments, where the value function is identified by an outcome-inducing bridge function. We estimate the bridge function by solving the conditional moment restriction problem using the minimax optimization in \citet{mastouri2021proximal}. %
We later obtain the optimal treatment regime by directly maximizing the empirical value function within a restricted class of IFTRs. To the best of our knowledge, this is the first article that provide a solution for deriving ITR for functional treatments in the presence of unmeasured confounding.
The performance of our method is demonstrated by a simulation study. We also apply our estimation method to the NHANES 2013-2014 accelerometry data and provide individualized PA distribution recommendations to participants based on their characteristics for the best of the Triglyceride-Glucose index. %

The main contribution of this paper is twofold. First, 
this paper makes the first attempt to study ITR for functional treatments in the presence of unmeasured confounding. The proposed ITR learning algorithm is flexible to learn both linear and nonlinear IFTRs. 
Moreover, we provide the first analysis of recommending optimal PA distributions for the best of the Triglyceride-Glucose index via NHANES data in the presence of possible unmeasured confounding. Our recommendation generally suggests participants increase their PA, while
specific recommendations may vary over covariate subgroups.
Some of the recommendations are aligned with the current PA
guidelines.

The rest of the paper proceeds as follows. In Section \ref{sec: IFTR ident}, we provide the IFTR formulation and identification result under the proximal causal inference framework.
Based on the identification result, we develop the IFTR learning method in Section \ref{sec: policy_learn}. In Section \ref{sec: IFTR/simu}, we present a simulation study to evaluate the performance of our proposed methods. The analysis of NHANES data is in Section \ref{sec: IFTR/realdata}. The conclusion and discussion are given in Section \ref{sec: IFTR/diss}. The code for this paper is publicly available on GitHub via \url{https://github.com/zlong66/IFTR}.
\vspace{-1em}

\section{IFTR Identification} \label{sec: IFTR ident}

\subsection{Problem setup}
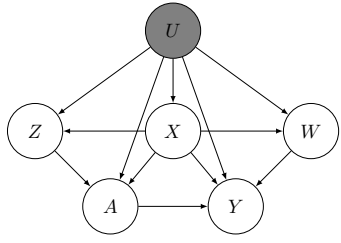
\begin{figure}
	\centering
	\vfill
	\resizebox{130pt}{!}{%
		\begin{tikzpicture}[state/.style={circle, draw, minimum size=1.1cm}]
  \def\Ax{0}
  \def\Ay{0}
  \def\offset{2.5}
  \def\Bx{\Ax+5}
  \def\By{\Ay}
  \node[state,shape=circle,draw=black] (Z) at (\Bx-4,\Ay+1.5) {$Z$};
  \node[state,shape=circle,draw=black] (Y) at (\Bx,\By) {$Y$};
  \node[state,shape=circle,draw=black] (A) at (\Bx-2.5,\By) {$A$};
  \node[state,shape=circle,draw=black] (X) at (\Bx-1.25,\By+1.5) {$X$};
    \node[state,shape=circle,draw=black, fill=gray] (U) at (\Bx-1.25,\By+3.5) {$U$};
  \node[state,shape=circle,draw=black] (W) at (\Bx+1.5,\Ay+1.5) {$W$};

  \draw [-latex] (X) to [bend left=0] (W);
  \draw [-latex] (A) to [bend left=0] (Y);
  \draw [-latex] (X) to [bend left=0] (A);
  \draw [-latex] (X) to [bend left=0] (Z);
  \draw [-latex] (X) to [bend left=0] (Y);
  \draw [-latex] (Z) to [bend left=0] (A);
  \draw [-latex] (W) to [bend left=0] (Y);

  \draw [-latex] (U) to [bend left=0] (A);
  \draw [-latex] (U) to [bend left=0] (Z);
  \draw [-latex] (U) to [bend left=0] (Y);
  \draw [-latex] (U) to [bend left=0] (W);
  \draw [-latex] (U) to [bend left=0] (X);

\end{tikzpicture}
	}
	\vfill
	\caption{A causal directed acyclic graph that illustrates variables under the
    proximal causal inference framework.
 }
	\label{fig:sra dag2}
\end{figure}

Let $Y$ denote the observed continuous outcome, and we assume that larger values of $Y$ imply better results. We let $A(\cdot)$ be the treatment, which is a function belonging to space $\mathcal{A}$. Without loss of generality, we assume that the domain of the functional treatment is $[0,1]$. 
The unmeasured confounders are denoted as $U$. 
Suppose that we collect a set of proxy variables to account for unmeasured confounding and can accurately allocate them into three types of proxies $(W, X, Z)$, where $X$ are proxies that are common causes of treatment $A$ and outcome $Y$, $W$ represents the outcome-inducing proxies that are related to $A$ only through $(U,X)$, and $Z$ denotes the treatment-inducing proxies that are related to $Y$ only through $(U,A,X)$. 
An example of the causal model under consideration is illustrated by Figure \ref{fig:sra dag2}. Under this causal model, the assumption of unconfoundedness is no longer valid, since when conditioning on $X$, $Y(a)$ and $A$ may be associated through the path $A-U-Y$.

Under the potential outcome framework, we denote $Y(a)$ as the potential outcome when treatment $A = a$, where $a$ is a function in $\mathcal{A}$. We adopt the standard consistency assumption $Y(A)=Y$, which connects the observed outcome $Y$ to the potential outcome $Y(a)$ when $A=a$. An individualized functional treatment regime (IFTR) $d$ is a mapping from the observable covariate space %
into a treatment space $\mathcal{A}$. The value function is the expected potential outcomes had the IFTR $d$ been implemented, or specifically,
   $ V(d)=E\{Y(d)\}$, %
where $Y(d)$ is the potential outcome if the IFTR $d$ were given.
 The optimal treatment regime $d_{opt} \in \argmax_{d \in {D}}E\{Y(d)\}$ is the regime that maximize the value function within an IFTR class ${D}$.

\subsection{Value function identification}\label{sec:value iden}

In the presence of unmeasured confounders, the value function identification results under the proximal causal inference framework have been developed by  \citet{qi2024proximal}. Their identification results mainly focus on binary treatments, and here we extend their results to functional treatments. We first list the basic assumptions needed for the identification result. 

\begin{assumption}[Assumptions for proxies]\label{assum2:w&z}
$$W(a,z)=W, \ Y(a,z)=Y(a) \text{ for all a and z almost surely.}$$
    
\end{assumption}

\begin{assumption}[Latent unconfoundedness]\label{assum2:unconf}
    $(Z,A) \perp (Y(a),W)|(U,X) \text{ for all a.}$
\end{assumption}

Assumption \ref{assum2:w&z} states that $A$ and $Z$ do not have a causal effect on $W$ and that $Z$ does not have a direct effect on $Y$ when intervening with $A$ \citep{tchetgen2024introduction}. This assumption will hold if $Z$ and $W$ are correctly specified as treatment-inducing and outcome-inducing proxies, respectively. Assumption \ref{assum2:unconf} is the generalization of the common unconfoundedness assumption, i.e., $Y(a) \perp A|X$ for all $a$. The latent unconfoundedness allows the existence of $U$, and will generally hold if $U$ is sufficient enough to include all the confounding between $(Z, A)$ and $(Y(a), W)$ not included in $X$. This assumption basically states that conditioning on $U, X$, the causal effect of $(Z,A)$ on $(Y(a), W)$ is identifiable. 

\begin{assumption}[Completeness]\label{assum2:complete}
    \begin{enumerate}
    \item[1)] For any a,x, if $E\{g(U)|Z,A=a, X=x\}=0$ almost surely, then $g(U)=0$ almost surely.
    \item[2)] For any a,x, if $E\{g(Z)|W,A=a, X=x\}=0$ almost surely, then $g(Z)=0$ almost surely.
\end{enumerate}
\end{assumption}

\begin{assumption}\label{assum2:h existence}
    There exists an outcome confounding bridge function $h(w,a,x)$ that solves the following integral equation
\begin{equation}
\label{eqn: h exist}
    E[Y|Z,A,X]=\int h(w,A,X)dF(w|Z,A,X),
\end{equation}
almost surely.
\end{assumption}

 The completeness assumption is a commonly made condition for identification problems of nonparametric and semi-parametric models \citep{newey2003instrumental, chen2018optimal, blundell2007semi,an2012well, carroll2010identification,shiu2013identification}.
Eq. \eqref{eqn: h exist} in 
Assumption \ref{assum2:h existence} connects the observed/primary outcome with a transformation of the negative control outcome, and Assumption 4 simply states that this transformation function $h$ exists. Note that Eq. \eqref{eqn: h exist} is a Fredholm integral equation of the first kind, and the existence of its solution is ensured under Assumption \ref{assum2:complete} (2) and some regularity conditions.  

If all assumptions listed above hold, then we are capable to identify the value function based on the observed variables using the following theorem, even though there exist unmeasured confounders. 

\begin{theorem} \label{thm:ITRidentify}
Let ${D}$ be the class of IFTRs that map from $\mathcal{X \times Z}$ to $\mathcal{A}$. Under Assumptions \ref{assum2:w&z}-\ref{assum2:unconf}, \ref{assum2:complete}(1) and \ref{assum2:h existence}, for any $d \in D$, the value function $V(d)$ can be nonparametrically identified by
\begin{equation}
    V(d)=E\{h(W,d(X,Z),X)\}.
\end{equation}
\end{theorem}

Theorem \ref{thm:ITRidentify} indicates that the value function is identifiable over the class $D$ when unmeasured confounding is present, and can be estimated if $h$ is known. This theorem is an extension of Theorem 3.1 in \citet{qi2024proximal} to functional treatments. 
The proof of Theorem \ref{thm:ITRidentify} is given in Appendix \ref{A: Thm proof}.

Based on the value identification result in Theorem \ref{thm:ITRidentify}, the optimal treatment regime can then be identified by
\begin{equation}
    {d_{opt}} \in \argmax_{d \in {D}} \, E\{h(W, d(X, Z), X\}. 
\end{equation}

\section{IFTR learning}\label{sec: policy_learn}

In this section, we propose an IFTR learning method 
based on the identification result in Theorem \ref{thm:ITRidentify}.
We start with the estimation of the nuisance function $h$. Then, we develop methods to learn optimal linear and nonlinear IFTRs respectively.

Our first step is to obtain the nuisance function $h$ that solve the Eq. (\ref{eqn: h exist}). Note that the Eq. (\ref{eqn: h exist}) can be written as  
$$E\{Y-h(W,A,X)|Z,A,X\} = 0.$$ Then, solving $h$ is a conditional moment restriction problem, for which various methods have been proposed \citep[e.g.,][]{dikkala2020minimax, zhang2023instrumental}. We adopt the Proxy Maximum Moment Restriction (PMMR) method proposed by \citet{mastouri2021proximal} to estimate 
$h$. Specifically, we obtain $\hat{h}$ by solving
\begin{equation}
\label{eq2:hobj}
    \hat{h} = \underset{h \in \mathcal{H}}{\arg\min} \left[ \frac{1}{n^2}\sum_{i,j=1}^{n}(Y_i-h_i)(Y_j-h_j)k_{ij} + \lambda \Vert h\Vert_{\mathcal{H}}^2 \right],
\end{equation}
 where $\lambda>0$ is a hyperparameter, $\Vert h\Vert_{\mathcal{H}}$ is the norm of $h$ in space $\mathcal{H},$ and $h_i \coloneqq h(W_i,A_i,X_i)$,  $k_{ij}\coloneqq k_{\mathcal{F}}([Z_i,A_i,X_i],[Z_j,A_j,X_j])$\footnote{The space $\mathcal{F}$ is a reproducing kernel Hilbert space, and any $f \in \mathcal{F}$ is a function on $\mathcal{Z} \times \mathcal{A} \times \mathcal{X}$. The function $k_{\mathcal{F}}$ is the kernel function in $\mathcal{F}$.}. Assuming that $\mathcal{H}$ is a reproducing kernel Hilbert space, there is a closed-form solution for $\hat{h}$. See more details in \citet{mastouri2021proximal}. 

With the estimated nuisance function $\hat{h}$, the optimal treatment regime can be obtained by directly maximizing the empirical version of the value function within the class $\mathcal{D}$, i.e., we can estimate IFTR via
$\hat{d} \in \argmax_{d \in \mathcal{D}} {n}^{-1} \sum_{i=1}^n \hat{h}(W_i, d(X_i,Z_i), X_i).$

To ensure that the estimated IFTR is a smooth function on its domain $[0, 1]$, we further penalize the roughness of the treatment regime. Consequently, the final optimal IFTR is obtained by solving the following optimization problem
\begin{equation}\label{eqn: find_d}
\begin{aligned}
    \hat{d} \in \argmax_{d \in {D}} \,
    \Bigg[ & n^{-1} \sum_{i=1}^n \hat{h}\{W_i, d(X_i, Z_i), X_i\}  - \rho \cdot E_{(X,Z)}\left\{\int_0^1 \left[d^{(2)}(X,Z)(t)\right]^2 dt \right\}
    \Bigg].
\end{aligned}
\end{equation}
where $ d^{(2)}(x,z)(t)$ is the second derivative of the IFTR with respect to $t$, 
and $\rho>0$ is a smoothing parameter. 

The treatment regime maps from $\mathcal{X \times Z}$ to the treatment space $\mathcal{A}$. In the following section, we consider two functional classes for linear and nonlinear treatment regimes, respectively, and present the algorithms for learning IFTRs in these two classes.

\subsection{Linear IFTR}\label{sec: lin IFTR}

Let $\mathbf{v} = ({x}^\top,{z}^\top)^\top$. We consider the class of linear IFTRs: 
\begin{equation*}
\begin{aligned}
\tilde{D}_l
= \Big\{
\tilde{d}(\mathbf{v}) :\ 
& \tilde{d}(\mathbf{v})(t)
= \beta_0(t) + \sum_{k=1}^{q} \beta_{k}(t) v_k,  \text{all } \beta_{k} \text{ are twice differentiable in } t
\Big\}.
\end{aligned}
\end{equation*}
where $t \in [0, 1]$, $q$ is the dimension of $({x},z)$.
Suppose that we can approximate $\beta_k(t) \approx \sum_{j=1}^J b_{k,j}\theta_j(t)$ for every $k$ where $\{\theta_{j}\}_{j=1}^J$ are cubic B-spline basis functions. 
Then, for each $d(\mathbf{v}) \in \tilde{D}_l$, we have
$d(\mathbf{v})({t}) \approx \mathbf{\tilde{v}^\top B \boldsymbol{\theta}}({t})^\top$, where $\mathbf{\tilde{v}}= (1,{x^\top, z^\top})^\top$,
$$\mathbf{B} = \begin{bmatrix}
   b_{0,1} & \cdots &  b_{0,J}\\
   \vdots & \ddots & \vdots \\
   b_{q,1} & \ldots & b_{q,J}
\end{bmatrix},$$and  $${\boldsymbol{\theta}({t})} = \begin{bmatrix}
\theta_1({t}) & \theta_2({t}) & \cdots & \theta_J({t}) 
\end{bmatrix}.$$ 
Similarly, we can approximate the penalty term in \eqref{eqn: find_d} as follows 
\begin{equation*}
    \begin{aligned}
        E_{\mathbf{V}} \left(\int_0^1 d^{(2)}(\mathbf{V})(t)^2 dt \right) & \approx E_V \left( \int (\mathbf{\tilde{V}^\top B \boldsymbol{\theta}^{(2)}}(t))^2dt\right)\\
        & =  tr \left(\mathbf{BRB^\top} E_V(\boldsymbol{\tilde{V}\tilde{V}^\top})\right),
    \end{aligned}
\end{equation*}
where $\mathbf{V}=(\mathbf{X^\top,Z^\top})^\top,$
$\boldsymbol{\theta^{(2)}}(t) = \begin{bmatrix}
\theta_1^{(2)}({t}), \theta_2^{(2)}({t}) , \cdots, \theta_J^{(2)}({t})
\end{bmatrix}$
and $\mathbf{R} = \int \boldsymbol{\theta^{(2)}}(t)^\top \boldsymbol{\theta^{(2)}}(t) dt$. 

In the optimization process, 
we approximate $E(\boldsymbol{\tilde{V}\tilde{V}^\top})$ by ${\hat{\boldsymbol{\Sigma}}} = {n}^{-1}\sum_i\boldsymbol{\tilde{V}_i\tilde{V}_i}^\top $, where $\mathbf{\tilde{V}}_i = (1, {X_i^\top,Z_i^\top})^\top$. 
The optimization problem in (\ref{eqn: find_d}) can then be 
transformed to 
the optimization of the parameter matrix $\mathbf{B}$, which is
\begin{equation}\label{eqn: find_d B l}
    \hat{\mathbf{B}} \in \argmax_{\mathbf{B}} \, \left[ n^{-1} \sum_{i=1}^n \hat{h}\{W_i, \boldsymbol{\tilde{V}_i^\top B\theta^\top}, X_i\}-\rho  \cdot tr\left(\boldsymbol{BRB^\top} \boldsymbol{\hat{\Sigma}}\right) \right].
\end{equation} 
The optimal IFTR is then 
obtained by 
$$\hat{d}(\mathbf{v})(t) = 
\mathbf{v^\top \hat{B}\boldsymbol{\theta}}({t})^\top.$$

\subsection{Nonlinear IFTR}\label{sec: nl IFTR}

The linear regime can be restrictive when the optimal treatment regime has a more complex relationship with covariates. 
Therefore, we consider another class for the treatment regime, which does not assume the linearity between the regime and the covariates and provides greater flexibility. 
Suppose that each regime $d: \mathbf{v} \rightarrow a(t) $, belongs to the following class of nonlinear IFTRs:

\begin{equation*}
\begin{aligned}
\tilde{D}_{nl}
= \Big\{
\tilde{d}(\mathbf{v}) :\ 
& \tilde{d}(\mathbf{v})(t) = f(\mathbf{v}, t),  f \text{ is twice differentiable in both } \mathbf{v} \text{ and } t
\Big\}.
\end{aligned}
\end{equation*}

for $t \in [0, 1]$. Suppose that $f$
can be well approximated by tensor product splines, that is, $$f(\mathbf{v},t) \approx  \sum_{j_0=1}^{N_0} \sum_{j_1=1}^{N_1}\cdots
\sum_{j_q=1}^{N_q}
b_{j_0, j_1, \ldots, j_q} \left\{ \theta_{0,j_0} (t)\prod_{k=1}^q \theta_{k,j_k}(v_k) \right\},$$
 where $\theta_{k,j_k}$ are cubic B-spline basis functions for $v_k$, and $\theta_{0,j_0}$ are cubic B-spline basis functions for $t$.
 Then for each $d(\mathbf{v}) \in \tilde{{D}}_{nl}$, we have 
\begin{align*}
    \tilde{d}(\mathbf{v})(t) 
    & = \sum_{j_0=1}^{N_0} \left\{\sum_{j_1=1}^{N_1}\cdots
\sum_{j_q=1}^{N_q}
b_{j_0, j_1, \ldots, j_q}\prod_{k=1}^q \theta_{k,j_k}(v_k)\right\}  \theta_{0,j_0}(t) = \boldsymbol{\psi(v) B\theta_0}(t)^\top,
\end{align*}
\noindent where $\boldsymbol{\psi(v)}$ is defined as  
$$\boldsymbol{\psi(v)}=\begin{bmatrix}
    \boldsymbol{\theta_1}(v_{1}) \otimes \boldsymbol{\theta_2}(v_{2}) \otimes \cdots \otimes \boldsymbol{\theta_q}(v_{q})
\end{bmatrix}$$ with $\boldsymbol{\theta_k}(v_k) = \left[\theta_{k,1}(v_k), ..., \theta_{k,N_k}(v_k)\right]$ and $\boldsymbol{\theta_0}(t) = \left[\theta_{0,1}(t), ..., \theta_{0,N_0}(t)\right]$ be the set of B-spline basis functions for $v_k$ and $t$, respectively, and $\otimes$ is the Kronecker product. The coefficient matrix is defined as $$\mathbf{B} = 
\begin{bmatrix}
    b_{1,1,...,1} & \cdots & b_{N_0,1,...1} \\
    \vdots & \ddots & \vdots \\
    b_{1,N_1,...,N_q} & \cdots & b_{N_0,N_1,...,N_q}    
\end{bmatrix}$$ with the $i$th matrix column arranged as {$(b_{i11...11},...,b_{i11...1N_q}, b_{i11...21}, ..., b_{i11...2N_q}, ..., b_{i11...N_{q-1}N_q}, ..., b_{iN_1...N_q})$}. 
We solve the optimization problem for the coefficient matrix $\mathbf{B}$
\begin{equation}\label{eqn: find_d B nl}
 \begin{aligned}
    \mathbf{\hat{{B}}} \in \argmax_{\mathbf{B}} \, \Big[n^{-1} \sum_{i=1}^n \hat{h}\{W_i, \boldsymbol{\psi(V_i) B\theta_0^\top}, X_i\} 
     -\rho \cdot \operatorname{tr}\left(\mathbf{BRB^\top} \boldsymbol{\hat{\Sigma}_{{\psi}}}\right) \Big],
\end{aligned}
\end{equation} 
where $\mathbf{V_i} = {(X_i^\top,Z_i^\top)}^\top$ and  $\boldsymbol{\hat{\Sigma}_{\psi}} = {n}^{-1}\sum_i \boldsymbol{\psi(V_i)^\top} \boldsymbol{\psi(V_i)}$. The estimated optimal IFTR is therefore $$\hat{d}(\mathbf{v})({t}) = \boldsymbol{\psi(v) \mathbf{\hat{B}}\theta_0}(t)^\top.$$

\subsection{Numerical optimizations and cross-fitting}

Note that the objective functions for both linear and nonlinear IFTRs, i.e., Eq. (\ref{eqn: find_d B l}) and (\ref{eqn: find_d B nl}), are not convex with respect to $\mathbf{B}$, 
but we can use the L-BFGS optimizer
to find the local maximum $\mathbf{\hat{{B}}}$. When optimizing linear IFTRs, we choose the initial value of $\mathbf{B}$ as the least squares estimator to predict $A$ with $\mathbf{V}$ as predictors. Details of computing the initial values are given in Appendix \ref{A: initial}. %
For nonlinear IFTR learning, 
we use multiple random restarts and select the best runs.

We adopt the cross-fitting procedure \citep[e.g.,][]{schick1986asymptotically, chernozhukov2018double, zhao2025efficient} 
for IFTR learning. 
First, we divide the data into $L$ folds. 
For each $l=1,...,L$, we estimate the nuisance function $\hat{h}^{(-l)}$ using data without the $l$th fold and obtain an estimate of the value function $\hat{V}^{(l)}(\mathbf{B})$ using the $l$th fold and estimated nuisance function $\hat{h}^{(-l)}$. We aggregate values from $L$ folds and obtain the cross-fitted value  $\hat{V}_{CF}(\mathbf{B})= {L}^{-1}\sum_{l=1}^L \hat{V}^{(l)}(\mathbf{B})$. Then, the optimal IFTR is estimated by $\hat{\mathbf{B}} = \argmax_{\mathbf{B}} \hat{V}_{CF}(\mathbf{B})+\rho\cdot PEN$, where $PEN$ is defined as $tr(\boldsymbol{BRB^\top\Sigma})$ for linear IFTR learning and as $tr(\boldsymbol{BRB^\top\Sigma_{\psi}})$ for nonlinear IFTR learning.

\section{Simulation Study} \label{sec: IFTR/simu}

We evaluate the performance of the proposed method via a simulation study. 

\subsection{Data generation}
The simulated datasets are generated using the following 
process. 
\begin{align*}
    & U \sim Unif(0,4), \\
    & X=[X_1,X_2]^\top; X \sim N([1, 1]^{\top},  \Sigma_X), \text{where } \Sigma_X = 
    \begin{bmatrix}
    1 &0.6 \\
    0.6 &1
    \end{bmatrix},
     \\
    & A := 0.25U\sin(4\pi t) 
    + 0.25X_1\sin(2\pi t) 
    + 0.25X_2cos(2\pi t) + \epsilon cos(6 \pi t), 
    \text{ where } \epsilon \sim \text{Normal}(0,0.1^2)\\
    & Z := 1.5U + 2X_1 + 1.5X_2 + 6\int_0^1 \sin(2\pi t)A(t)dt + \text{Normal}(0,0.5^2), \\
    & W :=U + 2X_1 + 3X_2 + \text{Normal}(0,0.5^2), \\
    & {Y :=
    1.2U + 0.8X_1 + 0.8X_2 - 10\int_0^1 (A(t)-d^{opt}(t))^2 dt  + \text{Normal}(0,0.5^2)}.
 \end{align*}
Note that the conditional mean of $Y$ given $(U, A, X)$ is a nonlinear function of the treatment $A$ and $d^{opt}(t)$, which is a function of covariates $(U,X)$ and will be defined later. The conditional mean of $Y$ also depends on the unmeasured covariate $U$. In addition, the value function will be maximized if the treatment 
$d(t) = d^{opt}(t)$, and thus $d^{opt}(t)$ is the true optimal treatment regime. Let {$\beta_1(t) = 0.12B_1(t)+0.36B_2(t)-0.36B_3(t)+0.36B_4(t)+0.36B_5(t)-0.6B_6(t)+0.24B_7(t), \quad
    \beta_2(t) = 0.16B_1(t)+0.32B_2(t)+0.16B_3(t), \quad
    \beta_3(t) = 0.05B_3(t)+0.1B_4(t)+ 0.1B_5(t),$} where $B_i: i=1,...7$ are 
    cubic B-spline basis functions on $[0,1]$ with three equally spaces interior knots.
    
We consider two scenarios for $d^{opt}(t)$, where $d^{opt}(t)$ is a linear function of covariates. 
\begin{itemize}
    \item[(L1)]     %
    $d^{opt}(t) := X_1 \beta_1(t) + X_2\beta_2(t) +U \beta_3(t)$.
    \item[(L2)] $d^{opt}(t) := X_1 \beta_1(t) + X_2\beta_2(t)$.
\end{itemize}
In scenario L1, there is a linear relationship between $d^{opt}(t)$ and covariates $X_1, X_2, U$. In scenario L2, 
the true treatment regime only depends on the observed covariates. 

We also consider two scenarios where $d^{opt}(t)$ is a nonlinear function. Define $\tilde{\beta_1}(t) = 0.2B_1(t)+ 0.4B_2(t), \quad
    \tilde{\beta_2}(t) = 0.12B_3(t)+0.24B_4(t), \quad
    \tilde{\beta_3}(t) = -0.4B_5(t) + 0.4B_7(t)$. The true regime functions are 
\begin{itemize}
    \item[(N1)]
    $d^{opt}(t) :=X_1X_2\tilde{\beta_1}(t)+ UX_1\tilde{\beta_2}(t)+UX_2\tilde{\beta_3}(t)$.
    \item[(N2)] 
    $d^{opt}(t) :=X_1X_2\tilde{\beta_1}(t) + X_1\tilde{\beta_2}(t) + X_2\tilde{\beta_3}(x)$.
\end{itemize}
 Both N1 and N2 encode a nonlinear relationship between the $d^{opt}(t)$ and covariates, but $d^{opt}(t)$ in N1 relies on $U$ while $d^{opt}(t)$ in N2 does not.
 
In scenarios $L1$ and $L2$, we generated simulated datasets with sample sizes $ n=500$ and $1000$ per dataset, whereas in scenarios $N1$ and $N2$, the datasets have a larger sample size with $n=5,000$. For each scenario, we also generate a test dataset with $n_{test} = 10,000$, with which we obtain estimated values. The estimated values are calculated using the equation $\widehat{E}\{Y(\hat{d})\}= n^{-1} \sum_{i=1}^n(1.2U_i+0.8X_{1i} + 0.8X_{2i} - 10 \int_0^1 (\hat{d_i}(t)-d^{opt}_i(t))^2dt)$, where $\hat{d}$ is the estimated IFTR. We run 100 simulations for each scenario. 

\subsection{Estimators in comparison}
We compare our proposed method with three other estimators denoted by $\hat{d}_{ORC}$, $\hat{d}_1$ and $\hat{d}_2$ respectively. 

If $(U,X)$ were observable, then according to Assumption \ref{assum2:unconf}, we have
\begin{align*}
V(d) & = E\{Y(d)\} \\
& = E[E\{Y(d)|X,U\}] \\
& = E[E\{Y|A=d(X,U),X,U\}]\\
& = E[m(X,U,d(X,U))], 
\end{align*}
where $m$ is the conditional mean of $Y$ given $(U, A,X)$. The conditional mean model maps scalar covariates and functional treatment to a scalar outcome and can be approximated by regression methods that accommodate functional covariates. In our simulation, we use the kernel ridge regression (KRR) to estimate $m$, that is, $\hat{m}$ is obtained by: $$\hat{m} = \argmin_{m \in \mathcal{H}_{UAX}} \left\{n^{-1}\sum_{i=1}^{n}(Y_i - m(U_i, X_i, A_i))^2+ \lambda \Vert m \Vert_{\mathcal{H}_{UAX}}^2 \right\}.\footnote{$\mathcal{H}_{UAX}$ is a RKHS with kernel $k: (\mathcal{U\times A\times X})^2 \rightarrow \mathbb{R}$. According to the representation theorem, 
$\hat{m} = \sum_{i=1}^n \alpha'_i K([X_i,U_i,A_i], \cdot), \text{where } \mathbf{\alpha'} = (\mathbf{K}+ n\lambda \mathbf{I})^{-1}\mathbf{Y},$ and $\mathbf{K}$ is a $n \times n$ matrix with $k([U_i,X_i,A_i], [U_j,X_j,A_j])$ as its $(i,j)$th entry.}$$ 
 Unlike typical scalar-on-function regressions which assume linearity, KRR provides a flexible way to model the nonlinear relationship between predictors and the outcome. Let ${D}_1$ be a class of functions that map $\mathcal{U\times X}$ to $\mathcal{A}$, we then obtain the estimated oracle IFTR by solving 

\begin{equation*}
\begin{aligned}
\hat{d}_{ORC} \in \argmax_{d \in {D_{1}} } \,
\Bigg[ & n^{-1} \sum_{i=1}^n \hat{m}\{X_i, U_i, d(X_i, U_i)\} \\
& - \rho \cdot E_{(U,X)} \left\{\int d^{(2)}(U,X)(t)^2dt \right\}
\Bigg].
\end{aligned}
\end{equation*}

The objective function is similar to \eqref{eqn: find_d}. Thus, we can use the same process described in Section \ref{sec: policy_learn} to convert the optimization problem for $d_{ORC}$ to an optimization problem for $\mathbf{B}$.

We consider two other IFTR estimators that ignore the existence of unmeasured confounders. Let ${D}_1$ be a class of functions mapping $\mathcal{X}$ to $\mathcal{A}$. For the estimator $d_1(X) \in {D}_1$, we only use $X$ and $A$ to estimate the conditional mean regression, then we learn the optimal IFTR $\hat{d}_1$ only in terms of $X$.
For $d_2(X) \in {D}_2$, we use all observed variables $(X,W,Z)$ together with $A$ to estimate the conditional mean regression function, and then obtain the optimal regime $\hat{d}_2$ that depends only on $X$. All conditional mean regressions are fitted using KRR. 

\subsection{Implementation details}
In $L1$ and $L2$, the regimes are fitted as linear functions of covariates, and we use 7 cubic B-spline bases to expand the coefficient functions. A sensitivity analysis was also conducted using 15 cubic B-spline basis functions, and the results are in Appendix \ref{A: add simu}. 
For scenarios $N1$ and $N2$, the regimes are fitted using the nonlinear IFTR learning method, where we use four cubic B-spline bases for each covariate and for $t$.   
The smoothing parameter $\rho$ is selected by the three-fold cross-validation, and we set the number of folds for cross-fitting $J = 3$. In $N1$ and $N2$, the kernels are approximated by Nystroem method \citep{williams2000using} to accelerate computation. 

Algorithm \ref{algrm: IFTR} 
summarizes the IFTR learning with cross-fitting, where $PEN$ is defined as $tr(\boldsymbol{BRB^\top\Sigma})$ for linear IFTR learning or as $tr(\boldsymbol{BRB^\top\Sigma_{\psi}})$ for nonlinear IFTR learning.

       \begin{algorithm}[h]\caption{Algorithm for IFTR learning}
  \label{algrm: IFTR}
    \begin{algorithmic}[1]
        \State Input: train data $(X_i, W_i, Z_i, A_i, Y_i)_{i=1}^n$
        \State Standardize train data$(X_i, W_i, Z_i)_{i=1}^n$, and find the best $\lambda$ in $h$.
        \For{$\rho$ in a pre-specified collection with size $M$} 
        \For {$k=1, \dots, K$}
        \For{$l=1, \dots, L$} 
            \State Estimate $\hat{h}^{(-k,-l)}$ with data of indices $I^{(-k,-l)}$
        \EndFor
        \State Find $\hat{\mathbf{B}}^{(k)}$ maximizes averaged empirical value $L^{-1}\sum_{l=1}^L\hat{V}^{(l)}(\mathbf{B}, {\hat{h}^{(-k,-l)}}) - \rho \cdot PEN$
        \State Calculate $v_k = \hat{V}^{(k)}(\hat{\mathbf{B}}^{(k)},\hat{h}^{(-k)})$
        \EndFor
        \EndFor
        \State Find $\rho^*$ that maximizes averaged empirical value $K^{-1}\sum_{k=1}^{K}v_k$ among $M$ tuning parameters.
        
        \For{$l=1, \dots, L$} 
            \State Estimate $\hat{h}^{(-l)}$ with data of indices $I^{(-l)}$
        \EndFor
        \State Find $\hat{\mathbf{B}}$ by maximizing $L^{-1}\sum_{l=1}^L\hat{V}^{(l)}(\mathbf{B}, {\hat{h}^{(-l)}})-\rho \cdot PEN$ using penalty coefficient $\rho=\rho^*$
        \State \textbf{Output:} Estimator $\hat{\mathbf{B}}$ and penalty $\rho^*$
        \State The value function is calculated in test data as $\hat{V}^{test}(\hat{\mathbf{B}}, {\hat{h}^{train}})$
    \end{algorithmic}
    \end{algorithm}

\subsection{Simulation results}
\begin{figure*} \centering \subfloat[L1]{ \includegraphics[width=0.45\textwidth]{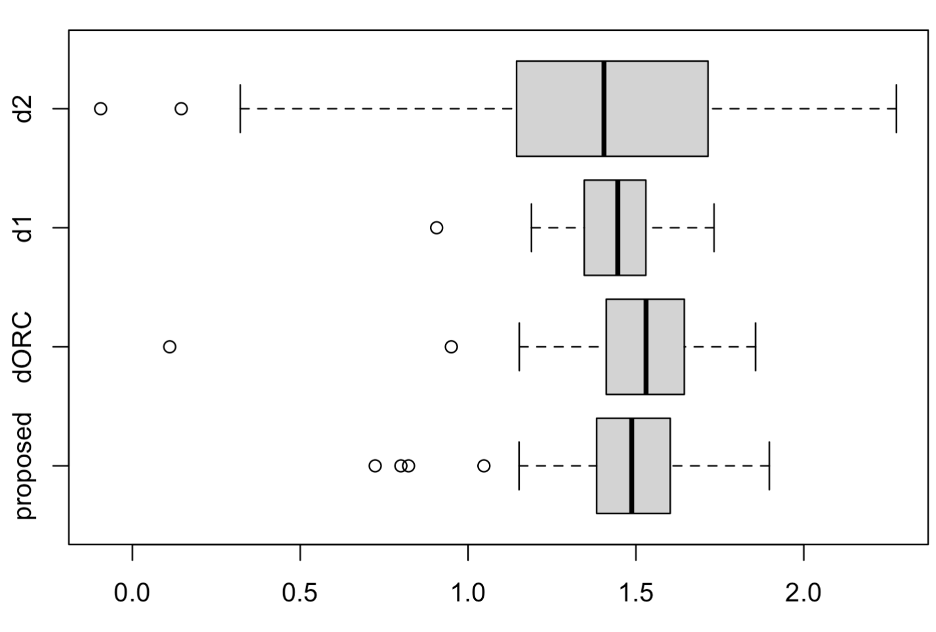} } \hfill \subfloat[L2]{ \includegraphics[width=0.45\textwidth]{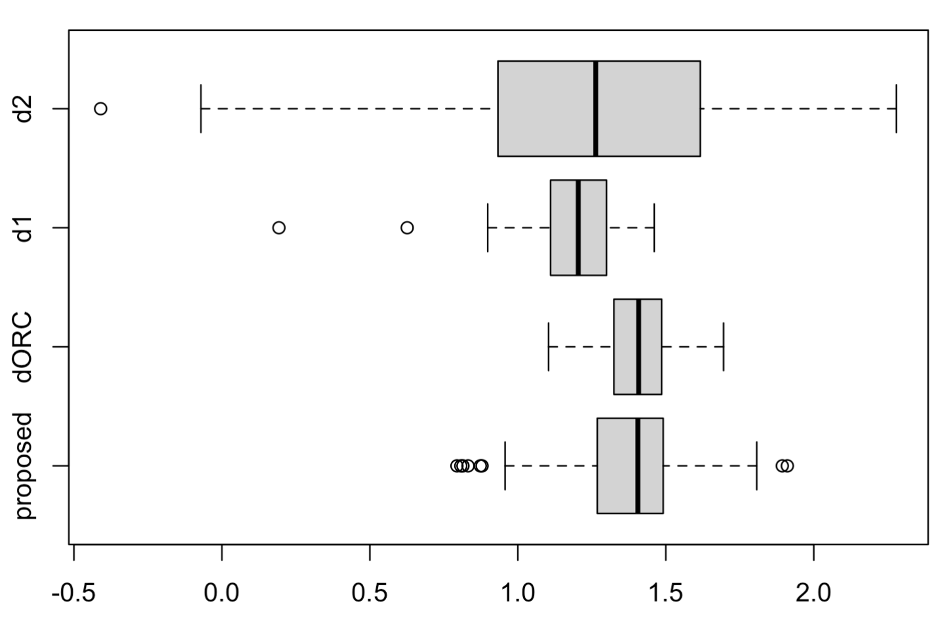} } \caption{The boxplots of values in scenarios L1 and L2 with $n=500$ and 7 basis functions.} \label{fig: IFTR bxplt1} \end{figure*}

\begin{figure*}[htbp]
\centering
\subfloat[L1]{
\includegraphics[width=0.45\textwidth]{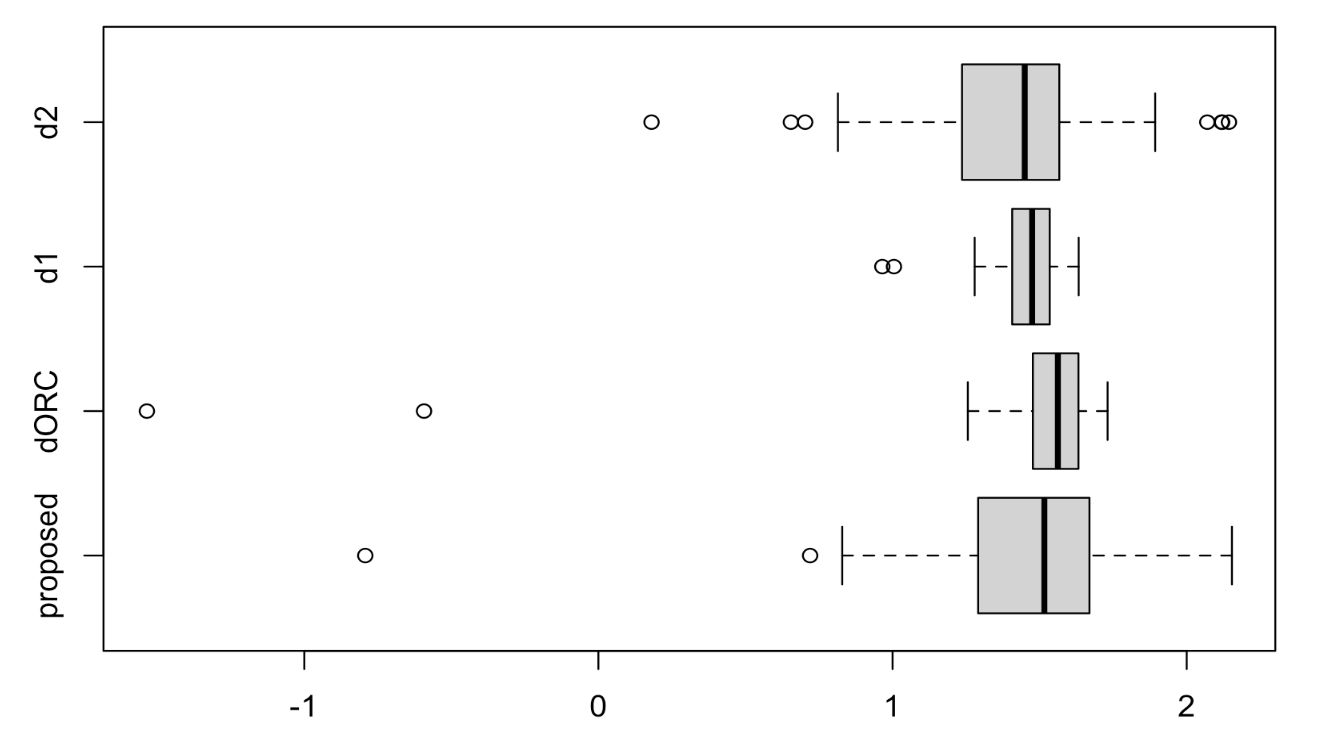}
}
\hfill
\subfloat[L2]{
\includegraphics[width=0.45\textwidth]{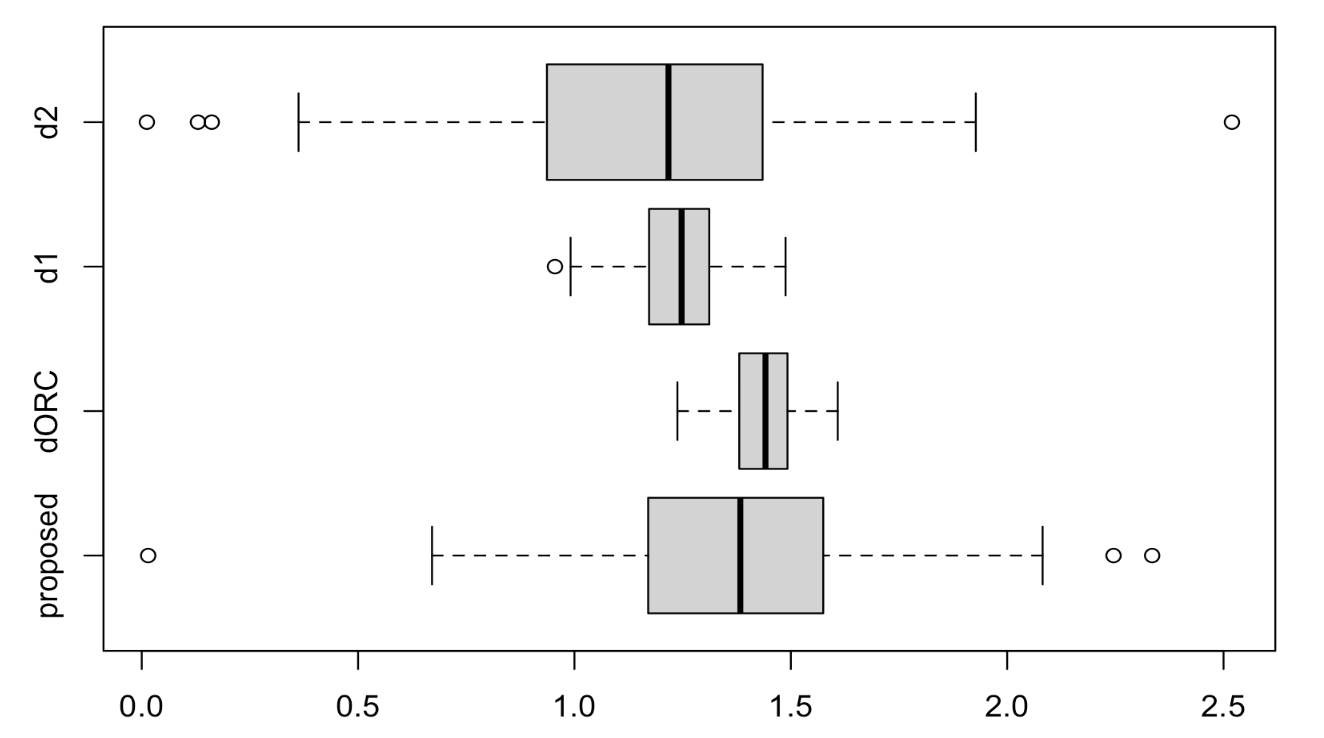}
}
\caption{The boxplots of values in scenarios L1 and L2 with $n=1,000$ and 7 basis functions.}
\label{fig: IFTR bxplt2}
\end{figure*}

\begin{figure*}[htbp] \centering \subfloat[N1]{ \includegraphics[width=0.45\textwidth]{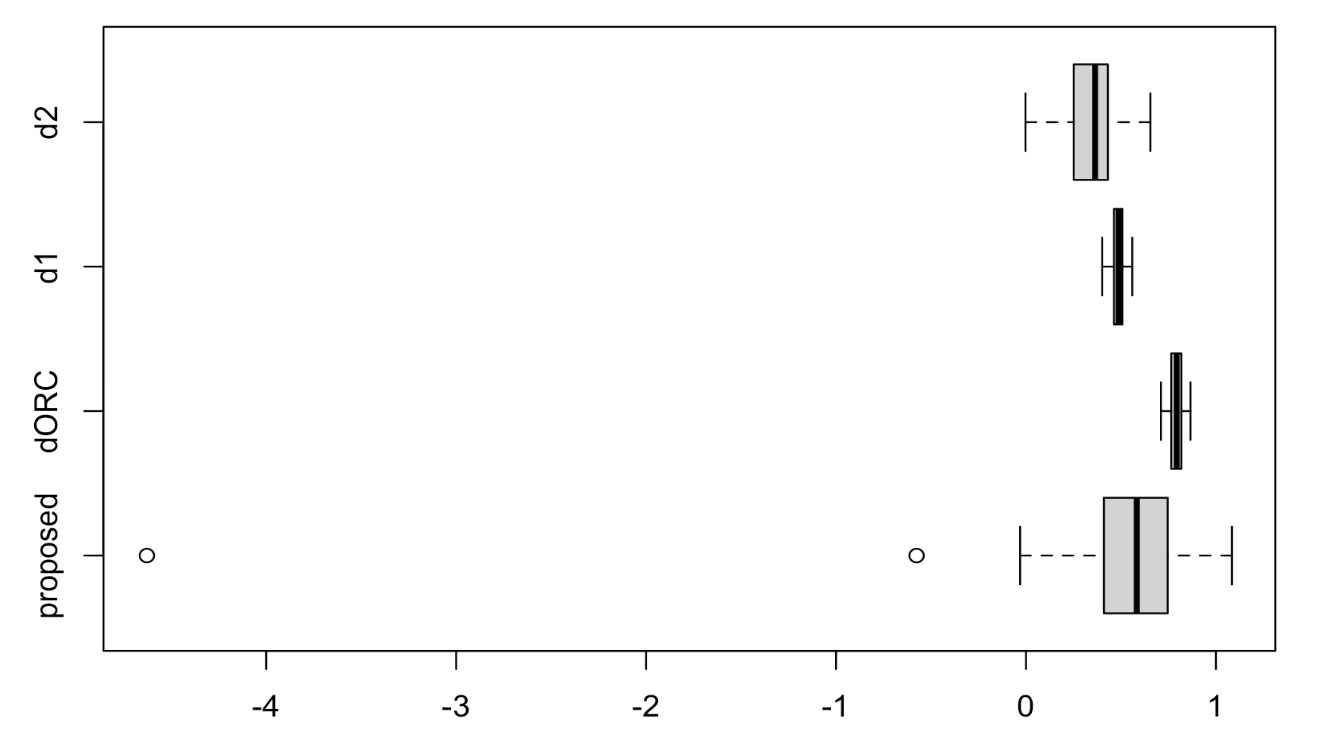} } \hfill \subfloat[N2]{ \includegraphics[width=0.45\textwidth]{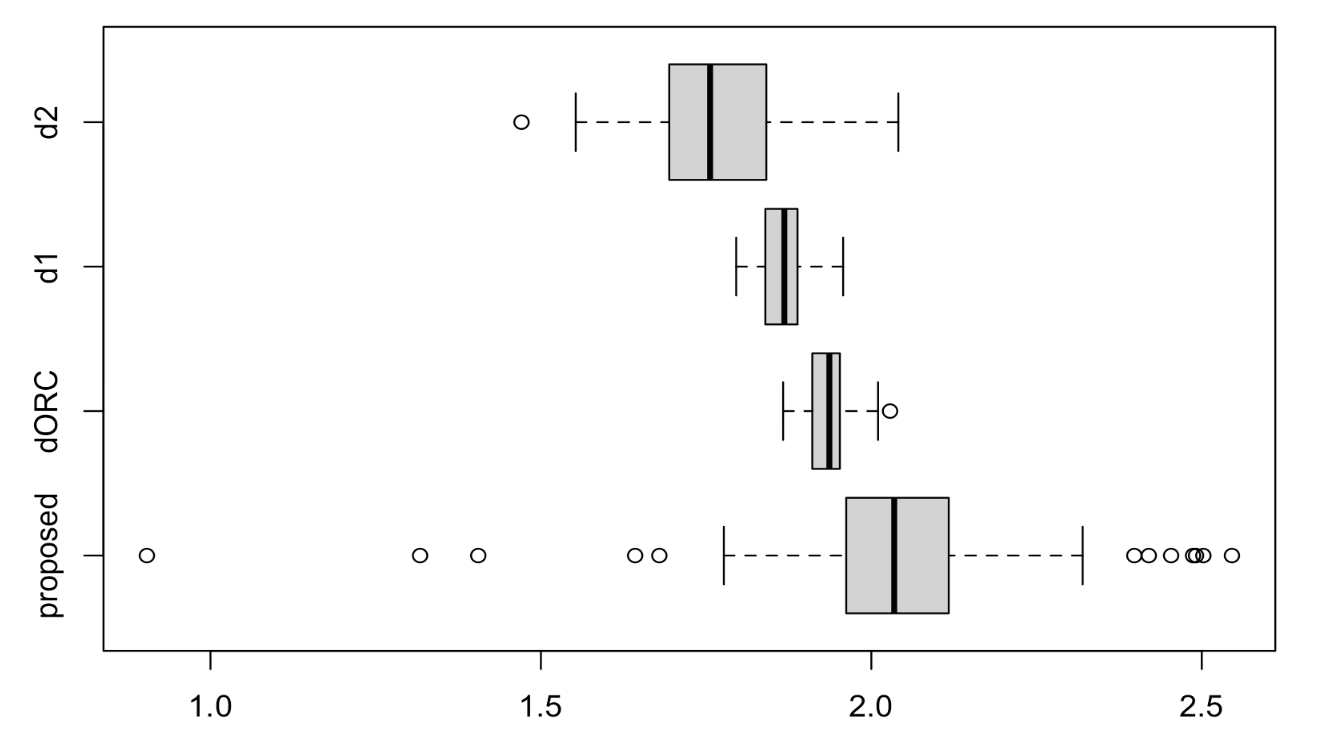} } \caption{The boxplots of values in scenarios N1 and N2 with $n=5,000$.} \label{fig: IFTR bxplt3} 
\end{figure*}

The boxplots of empirical values estimated using four methods under Scenarios L1 and L2 are shown in Figure \ref{fig: IFTR bxplt1} and \ref{fig: IFTR bxplt2}, and the results for Scenarios N1 and N2 are given in Figure \ref{fig: IFTR bxplt3}.
In both L1 and N1, the oracle method is the best among all methods, followed by the proposed method. It is reasonable since the oracle method incorporates the unmeasured covariate in the regime, while the proposed method does not. Regimes $\hat{d}_1$ and $\hat{d}_2$ perform worse than the proposed method and the oracle method because they fail to account for the unmeasured confounder $U$. %
In Scenarios L2 and N2, where the true regime does not contain $U$, we observe a similar pattern, 
except that the proposed method is more comparable with or even has a better performance than the oracle method.

\section{Real data analysis} \label{sec: IFTR/realdata}

\subsection{Data description}
 The National Health and Nutrition Examination Survey (NHANES) collects information from around 5,000 participants representing different populations in the US on their health and nutrition status. Typically, participants undergo an interview, body and health measurements, and laboratory tests. During the 2011-2014 cycle, participants were also asked to wear physical activity monitors (PAM) for seven consecutive days to collect their 24-hour movements. Every 1/80 second (80 Hz), the PAM collects the acceleration of the subject on three axes (x-, y-, z-axis). The raw PAM data were then summarized at the minute level using an open-source algorithm, specified in Monitor-Independent Movement Summary (MIMS) units. In addition to minute-level PA intensity measures, the data also contains a variable representing the predicted status, i.e., ``sleep", ``wake", ``non-wake", or ``unknown", for each minute.

 In this analysis, we use data collected during the 2013-2014 cycle. We focus on participants who wear PAM on the non-dominant wrist and have PA records for at least 3 valid weekdays and 1 valid weekend day. Following the criteria of \citet{to2022differences}, a valid day is a day with at least 1389 valid minutes, fewer than 72 minutes of non-wear time, and less than 17 hours of sleeping wear. 
 
  The health biomarker of interest in this analysis is Triglyceride-Glucose index ({TyG}). 
  TyG is a composite biomarker composed of fasting triglyceride and fasting glucose levels, 
  and it is calculated as 
 $\log(\text{Triglyceride} \times \text{Fasting Glucose}/2).$ In addition to being an accessible tool that measures insulin resistance and metabolic health of subjects, recent studies have shown its predictive power for the risk of diseases like type 2 diabetes, cardiovascular diseases, and non-alcoholic fatty liver disease among others \citep{tao2022triglyceride, nayak2024diagnostic}.
 The outcome $Y$ is defined as $1/\text{TyG}$ so that a larger value of $Y$ generally indicates better metabolic health and insulin sensitivity. 
 
 We identified 8 proxy variables, with two outcome-inducing variables ($W$) related to the family history of diabetes or heart attack, one treatment-inducing variable ($Z$) measuring the frequency of using sunscreen, and five common cause variables ($X$), including age, sex, race, body mass index and an indicator for cardiovascular symptoms. The cardiovascular symptoms indicator is set to 1 if the subject aged greater or equal to 40, and has chest pain or short breath, and 0 otherwise. Detailed information on the selected proxy is provided in Table \ref{tab:variables}.  
 After removing incomplete data from proxy variables and the outcome, the dataset contains 1044 participants. We further removed subjects identified as ``Other" race, as the number of subjects in the other races is relatively small in the dataset. After the data preprocessing, our final dataset has 1,009 participants.

We took the following steps to obtain the functional treatment $A$. 
First, based on nonzero MIMS measurements only, we obtained the density of log-transformed PA intensity ($MIMS\rightarrow log(MIMS+1)$ ) within a range $(0, 5)$ using the kernel density estimator. The bandwidth was selected by Silverman's rule of thumb method. We chose the upper bound $5$ because less than 1\% of the PA measurements exceeded this value. Then, to ensure that all PA density functions are strictly positive on $(0,5)$, each density was regularized by mixing it with a uniform distribution on $(0,5)$. The mixing coefficient was selected so that the resulting density was bounded below by 0.005 throughout the interval \citep{petersen2021wasserstein}. Since each kernel density is subject to the constraint that its integral must equal to one, we finally
applied the log-quantile density transformation 
to the kernel densities to obtain the ultimate functional treatments $A_i$ which are  unconstrained functions \citep{petersen2016functional}.

The goal of this analysis is to estimate an optimal treatment regime that assigns a distribution of physical activity to a participant based on his/her characteristics for the best of TyG. 
Due to the limited sample size, we focus on linear IFTRs for our proposed method as well as $\hat{d}_1$ and $\hat{d}_2$ that ignore unmeasured confounding.
When optimizing IFTRs as described in Section \ref{sec: lin IFTR}, 
we used 15 
cubic B-spline basis functions defined over the interval $[0,1]$ with equally spaced interior knots.

\subsection{Results}

First, we compute empirical values by three-fold cross-validation. 
Specifically, we divide the data into three folds, using two folds as the training data for the treatment regime estimation, and the remaining fold as the test data, with which we calculate the empirical value using the estimated regime. The process is then repeated three times until all folds have been used as test data once. The obtained empirical values are summarized in Table \ref{tab:real: values}. 
Table \ref{tab:real: values} shows that 
the proposed method has the highest empirical values uniformly and thus outperforms the other two methods, which ignore the existence of unmeasured confounders. %
\begin{table}[h!]
\centering
\caption{The empirical values obtained using three estimators.}
\label{tab:real: values}
\setlength{\tabcolsep}{12pt}
\begin{tabular}{lcccc}
\toprule
Methods & CV1 & CV2 & CV3 & Average \\
\midrule
$d_1$ & 0.130 & 0.135 & 0.130 & 0.132 \\
$d_2$ & 0.131 & 0.134 & 0.129 & 0.131 \\
Proposed & \textbf{0.147} & \textbf{0.148} & \textbf{0.149} & \textbf{0.148} \\
\bottomrule
\end{tabular}
\end{table}

Next we assess the difference between the recommended PA distributions by our proposed method and behavior PA distributions for all subjects as well as subgroups stratified by a few covariates. For ease of interpretations, after we obtain the optimal IFTR in the log-quantile density scale, we transform it to a quantile function.

\begin{figure}[h]
    \centering
    \includegraphics[width=0.45\textwidth]{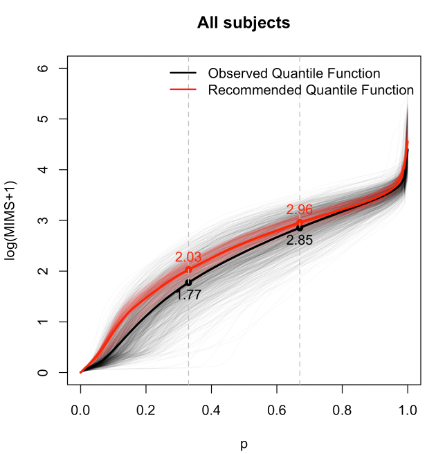}
    \caption{Comparison of the observed PA distributions and the PA distributions recommended by the estimated IFTR. The x-axis represents the percentile/quantile level, and the y-axis represents the corresponding quantile function values in the scale of $\log(\mathrm{MIMS}+1)$. 
    The transparent red curves represent the recommended PA distributions, while the transparent black curves represent the observed PA distributions. The bold red and black curves denote their cross-sectional means 
    respectively. The dashed vertical lines are located at $p = 0.33$ and $p = 0.67$. 
    }
    \label{fig: IFTR all}
\end{figure}

Figure \ref{fig: IFTR all} illustrates the recommended PA distributions by our proposed method against behavior PA distributions for all subjects. As shown in Figure \ref{fig: IFTR all}, the proposed method suggests people uniformly engage in more PA than their behaviors to improve their TyG.

\begin{figure*}[htbp]
\centering
\subfloat[]{
\includegraphics[width=0.45\textwidth]{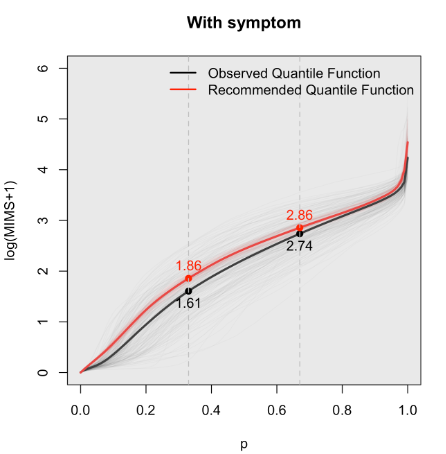}
}
\hfill
\subfloat[]{
\includegraphics[width=0.45\textwidth]{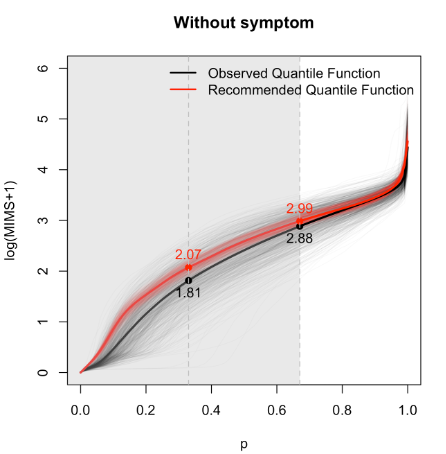}
}
\caption{Comparison of the observed and recommended PA distributions for participants with and without cardiovascular symptoms. See Figure~\ref{fig: IFTR all} for a description of the graphical elements.}
\label{fig: IFTR symptom}
\end{figure*}

\begin{figure*}[htbp]
\centering
\subfloat[]{
\includegraphics[width=0.31\textwidth]{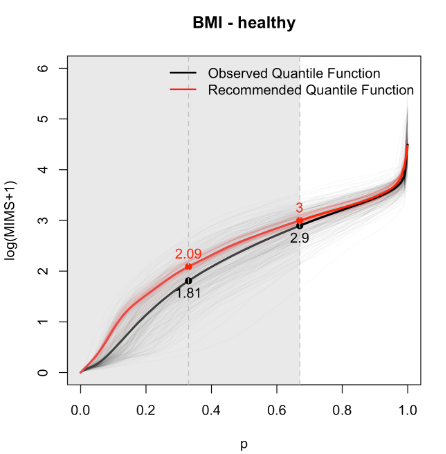}
}
\hfill
\subfloat[]{
\includegraphics[width=0.31\textwidth]{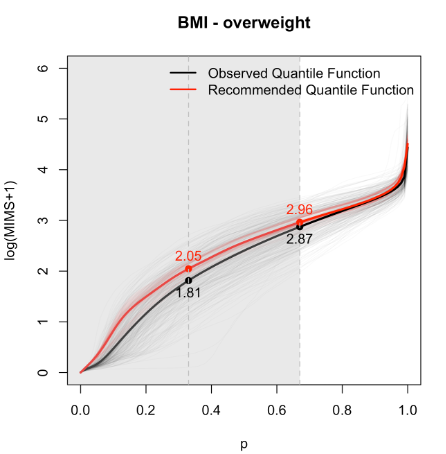}
}
\hfill
\subfloat[]{
\includegraphics[width=0.31\textwidth]{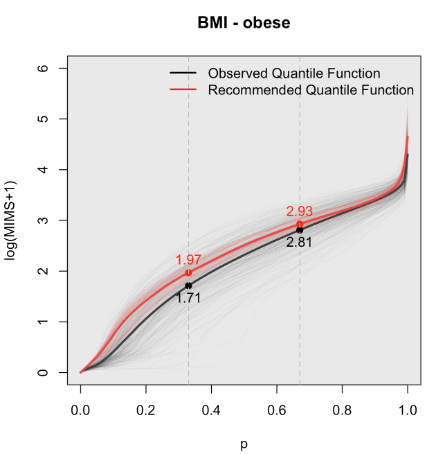}
}
\caption{Comparison of the observed and recommended PA distributions by BMI. See Figure~\ref{fig: IFTR all} for a description of the graphical elements.}
\label{fig: IFTR bmi}
\end{figure*}

\begin{figure*}[htbp]
\centering
\subfloat[]{
\includegraphics[width=0.45\textwidth]{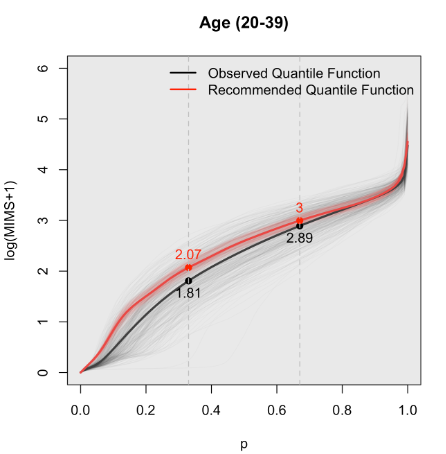}
}
\hfill
\subfloat[]{
\includegraphics[width=0.45\textwidth]{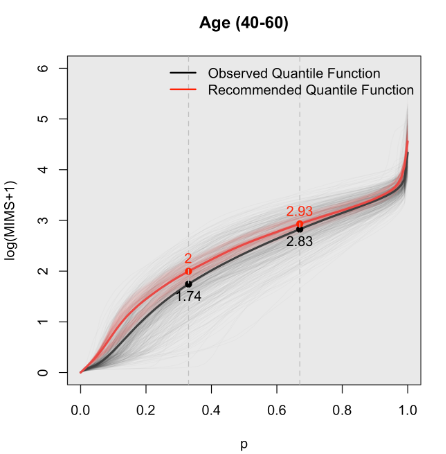}
}
\caption{Comparison of the observed and recommended PA distributions by age. See Figure~\ref{fig: IFTR all} for a description of the graphical elements.}
\label{fig: IFTR age}
\end{figure*}

\begin{figure*}[htbp]
\centering
\subfloat[]{
\includegraphics[width=0.45\textwidth]{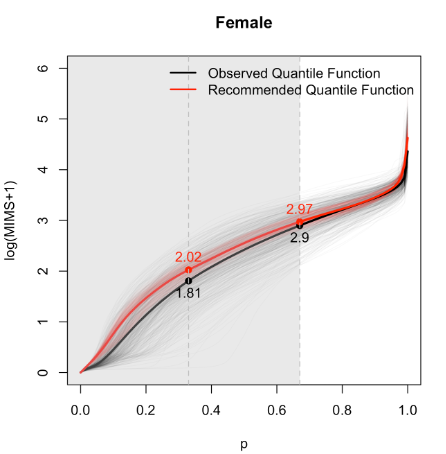}
}
\hfill
\subfloat[]{
\includegraphics[width=0.45\textwidth]{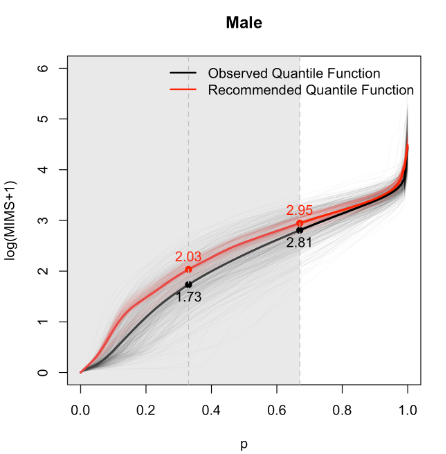}
}
\caption{Comparison of the observed and recommended PA distributions by sex. See Figure~\ref{fig: IFTR all} for a description of the graphical elements.}
\label{fig: IFTR gender}
\end{figure*}

We also conduct a subgroup analysis to examine how the recommended PA distributions vary according to important covariates. Here we only focus on subgroups of subjects stratified by having cardiovascular symptoms or not, body mass index (BMI), age, and sex respectively. 
Within each subgroup, 
we applied the interval-wise testing procedure for functional data \citep{pini2017interval} to test significance for the paired difference between the recommended and observed PA quantile distributions. The results are presented in Figures \ref{fig: IFTR symptom}--\ref{fig: IFTR gender}. In each figure, the shaded regions among $[0, 0.33]$, $[0.33,0.67]$, and $[0.67, 1]$ indicate quantile intervals where
the means of the recommended and observed PA quantile distributions differ
significantly, defined as regions where the adjusted p-value function was less than the interval-wise Type I error probability $0.05$. The three quantile intervals $[0, 0.33]$, $[0.33,0.67]$, and $[0.67, 1]$ may be understood to correspond to PA of lower, moderate, and higher intensities, respectively.

As shown in Figure \ref{fig: IFTR symptom} (a), the mean recommended PA distribution for participants with cardiovascular symptoms is significantly higher than the mean observed PA distribution across all three quantile intervals. In contrast, for participants without symptoms, significant differences are observed only in the first two quantile intervals as illustrated in Figure \ref{fig: IFTR symptom} (b). These findings suggest that the proposed method recommends increasing overall PA intensity for individuals with cardiovascular symptoms, whereas individuals without symptoms are mainly encouraged to increase PA at lower and moderate intensity levels. Furthermore, we observe that participants with cardiovascular symptoms are recommended a lower-intensity PA distribution than those without symptoms. This suggests that, although individuals with symptoms 
are suggested to increase their PA, their recommended intensity should not be as high as that for asymptomatic individuals to ensure safety \citep[e.g.,][]{whitfield2017applying}.

Similarly, for BMI subgroups, Figure \ref{fig: IFTR bmi} suggests obese participants improve their overall PA intensity, while recommend the healthy and overweight participants to mainly increase their PA intensity at lower and moderate intensity levels. In addition, the recommended PA levels for obese individuals are lower than their healthy or overweight counterparts. 
The results are aligned with the recommendations that individuals with obesity are advised to begin with moderate-intensity PA before progressing to vigorous-intensity PA \citep[e.g.,][]{pescatello2021EP}. 

As shown in Figure \ref{fig: IFTR age}, for age subgroups, the recommended PA distributions have significantly higher intensity than the observed PA distributions for both younger participants (aged 20--39) and older participants (aged 40--60). In addition, the proposed method recommends a slightly lower PA intensity for older participants than for younger participants. This echoes the recommendation that although the general adult guidelines also apply to older adults, older individuals should determine an appropriate PA level based on their fitness and functional abilities \citep[e.g.,][]{piercy2018physical}.

Regarding sex subgroups, Figure \ref{fig: IFTR gender} shows that, although the observed PA distributions among males have lower intensity than those among females, the proposed method recommends similar PA distributions for both sexes.

\section{Conclusion and Discussion} \label{sec: IFTR/diss}

In this paper, we make the first attempt to learn ITR for functional treatments under unmeasured confounding. Under the proximal causal inference framework, we establish the value function identification result based on which, together with penalized spline smoothing, we propose an algorithm to learn the optimal and smooth IFTR. 
Applying our proposed method to the NHANES accelerometry data, we obtain the first individualized PA distribution recommendation that optimizes the Triglyceride-Glucose index. It generally suggests participants increase their PA, while specific recommendations may vary over covariate subgroups. Some of the recommendations are aligned with the current PA guidelines. Our simulation study also shows that the proposed method has better performance than methods that do not account for unmeasured confounding.

When the number of covariates is large, nonlinear IFTRs may involve a substantial number of basis coefficients, and the sample size of data may not be sufficient for the coefficients optimization. Variable selection is therefore important for reducing complexity and improving the interpretability of the estimated regime. Future work may incorporate group LASSO or related penalization methods to select relevant covariates. Such methods may also be useful for linear IFTRs, since covariates included for nuisance function estimation are not necessarily relevant to the treatment regime.

For future work, we will study the theoretical properties of our
proposed method. Moreover, our proposed method is applicable to find individualized PA distribution recommendations for other health outcomes if suitable proxy variables can be identified. The number of variables used in the NHANES analysis is relatively small. Future work could include more covariates when recommending the individualized PA distribution, for example, taking into account exercise goals and more health conditions, including hypertension, diabetes, etc.

\subsection*{Acknowledgments}

Zhang's research is partially supported by the George Washington University Cross-Disciplinary Research Fund.

\subsection*{Conflicts of Interest}

The authors declare no conflicts of interest.

\subsection*{Data Availability Statement}

The data that support the findings of this study are openly available in NHANES at \url{https://wwwn.cdc.gov/nchs/nhanes/default.aspx}.

\bibliographystyle{apalike}
\bibliography{reference}

\clearpage
\appendix

\renewcommand{\thefigure}{\thesection\arabic{figure}}
\renewcommand{\thetable}{\thesection\arabic{table}}
\renewcommand{\theequation}{\thesection\arabic{equation}}

\section*{Appendix}
\addcontentsline{toc}{section}{Appendix}

The appendix includes the proof of Theorem \ref{thm:ITRidentify}, 
initialization for the 
linear IFTR optimization, additional simulation results for linear IFTR, and detailed information on the variables used in the real data analysis.

\section{Proof of Theorem \ref{thm:ITRidentify}} \label{A: Thm proof}

First, we show that
\begin{align*}
    E\{Y(d)|X,Z\} &= E[E\{Y(d)|X,U,Z\}|X,Z]\\
    &=E[E\{Y(d)|X,U\}|X,Z]\\
    &=E[E\{Y|X,U, A=d(X,Z)\}|X,Z]\\
    &=E[E\{h_0(W,d(X,Z),X)|X,U\}|X,Z]\\
    &=E[E\{h_0(W,d(X,Z),X)|X,U,Z\}|X,Z]\\
    &=E\{h(W,d(X,Z),X)|X,Z\}.
\end{align*}
The second, third and fifth equality are due to Assumption \ref{assum2:unconf}, and the fourth equality is shown by Theorem 1 in \citet{miao2018identifying} under Assumptions \ref{assum2:complete} (1) and \ref{assum2:h existence}.

Therefore, we have
\begin{equation*}
    \begin{aligned}
        V(d) &= E\{Y(d)\} \\
        & = E[E\{Y(d)|X,Z\}] \\
        & = E\{E[h\{W,d(X,Z),X\}|X,Z]\} \\
        & =E\{h(W,d(X,Z),X)\}.
    \end{aligned}
\end{equation*}

\section{Initialization for Linear IFTR} \label{A: initial}
    We assume $$ E(A(t)|V) = \beta_0(t)+\sum_{k=1}^{q} V_{k}\beta_k(t) \approx \sum_{k=0}^{q} \sum_{j=1}^{J}b_{k,j}V_k \theta_{j}(t) = \tilde{V}^\top\mathbf{ B }\boldsymbol{\theta}({t})^\top. $$ 
    Let $\mathbf{\tilde{V}}$ be the $n\times(q+1)$ matrix with  $\mathbf{\tilde{V}} =[\mathbf{1}, \mathbf{V_1}, \cdots,\mathbf{V_q}],$
    and $\mathbf{b} = vec(\mathbf{B^\top}) = [\mathbf{b_0^\top}, \mathbf{b_1}^\top,...,\mathbf{b_q}^\top].$ Let $\mathbf{A}$ and $\boldsymbol{\Theta(t)}$ be the matrices that evaluate $A(\cdot)$ and $\boldsymbol{\theta}(\cdot)$ at points $\mathbf{t} = (t_1,...,t_N)$, respectively. Then, we have  $$E(\mathbf{vec(A^{\top})}) = [\mathbf{\tilde{V}}  \otimes \boldsymbol{\Theta(t)}] \mathbf{b}.$$ The ordinary least squares (OLS) estimator \citep{ramsay2005functional} 
    $$\hat{b}_{ols} = \left[(\mathbf{\tilde{V}^\top \tilde{V}})\otimes\boldsymbol{(\Theta^\top \Theta)}\right]^{-1}(\boldsymbol{\tilde{V} \otimes \Theta})^{\top}vec(\mathbf{A^\top}),$$ is then obtained and used as the initial value for $\mathbf{b}$. We further transform it back to the matrix form to obtain $\mathbf{B_0}$, the initial values for $\mathbf{B}$.

\section{Additional simulation results}\label{A: add simu}

\setcounter{figure}{0}
\setcounter{table}{0}
\setcounter{equation}{0}

\begin{figure}[htbp]
\centering
\subfloat[L1]{
\includegraphics[width=0.45\textwidth]{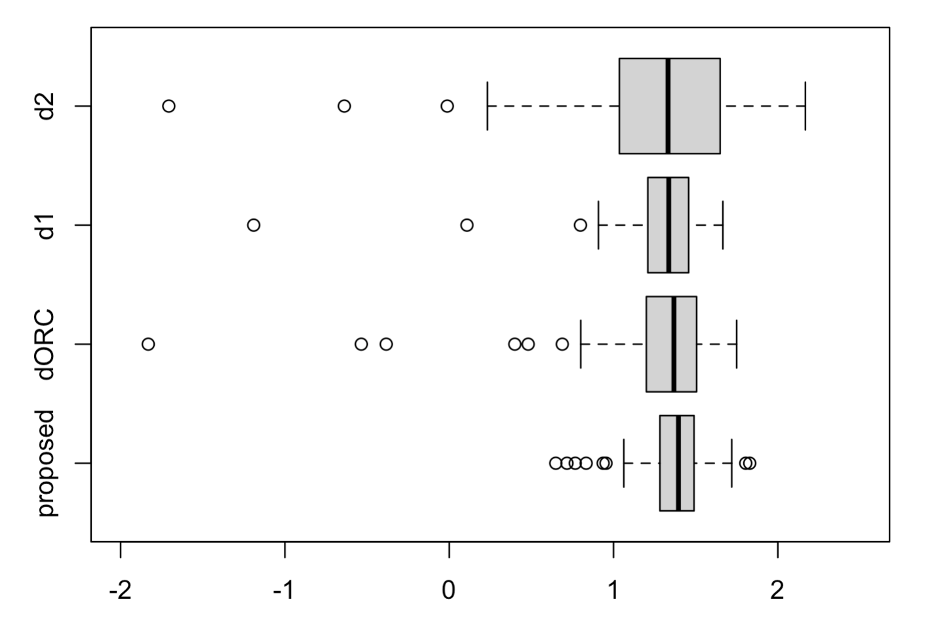}
}
\hfill
\subfloat[L2]{
\includegraphics[width=0.45\textwidth]{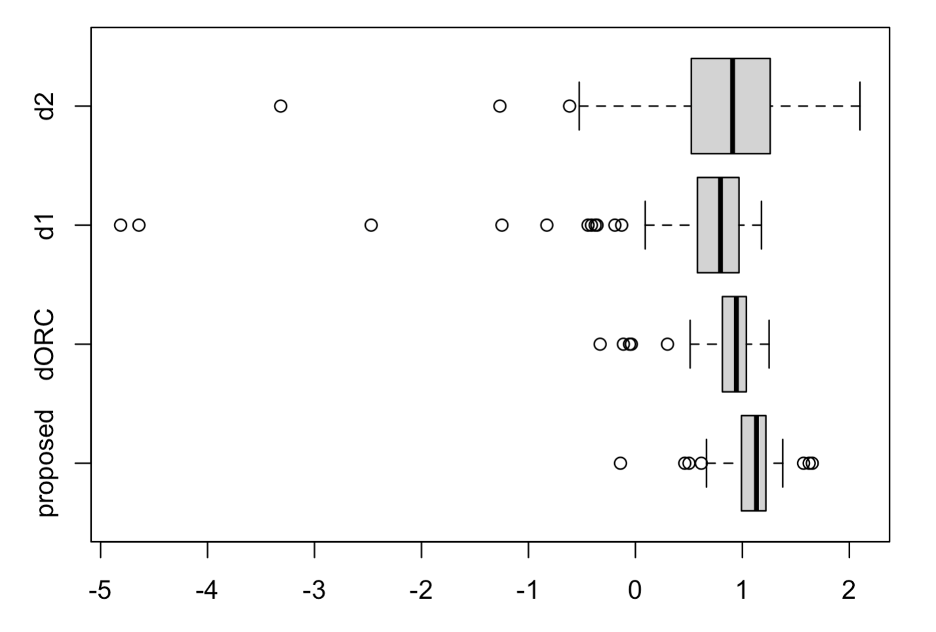}
}
\caption{The boxplots of values in scenarios L1 and L2 with $n=500$ and 15 basis functions.}
\label{fig: IFTR 500_15}
\end{figure}

\begin{figure}[htbp]
\centering
\subfloat[L1]{
\includegraphics[width=0.45\textwidth]{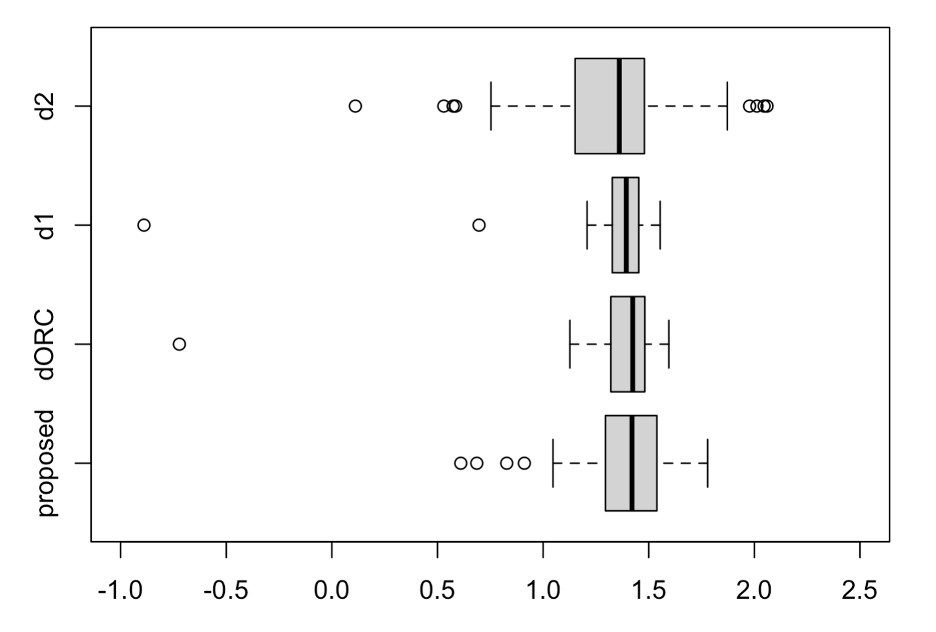}
}
\hfill
\subfloat[L2]{
\includegraphics[width=0.45\textwidth]{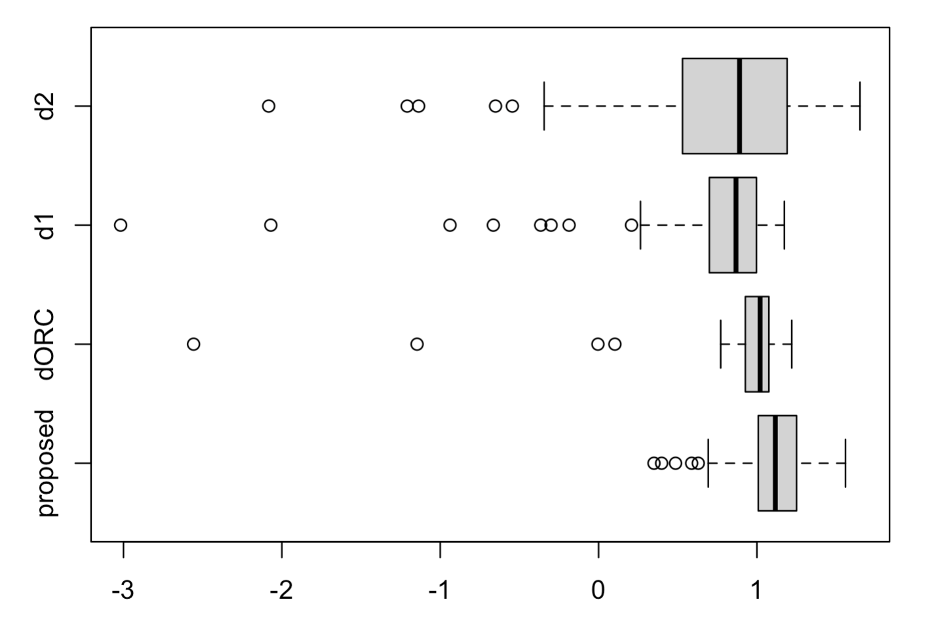}
}
\caption{The boxplots of values in scenarios L1 and L2 with $n=1,000$ and 15 basis functions.}
\label{fig: IFTR 1000_15}
\end{figure}

\begin{figure}[htbp]
\centering
\subfloat[L1 ($n=500$)]{
\includegraphics[width=0.45\textwidth]{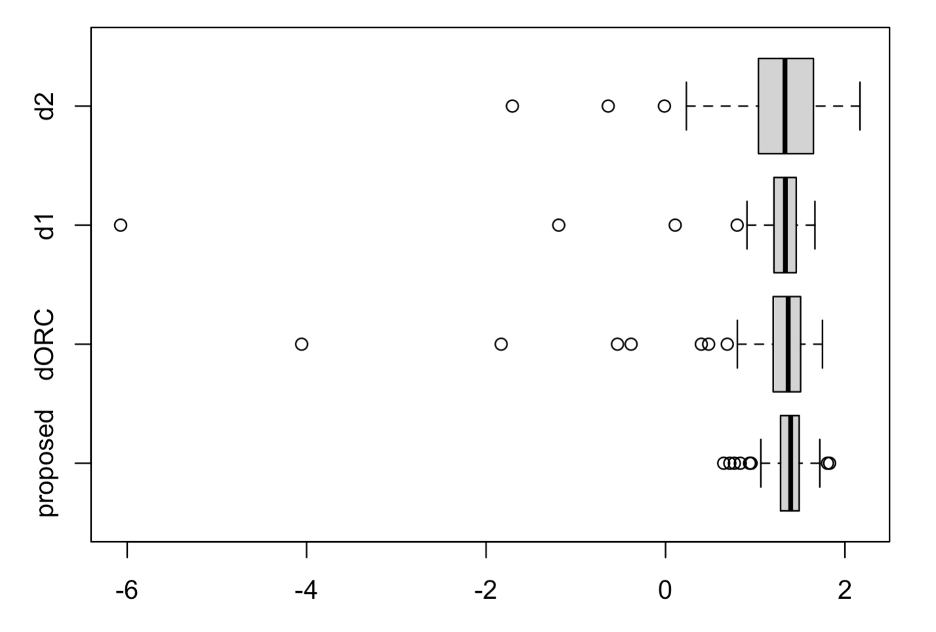}
}
\hfill
\subfloat[L1 ($n=1,000$)]{
\includegraphics[width=0.45\textwidth]{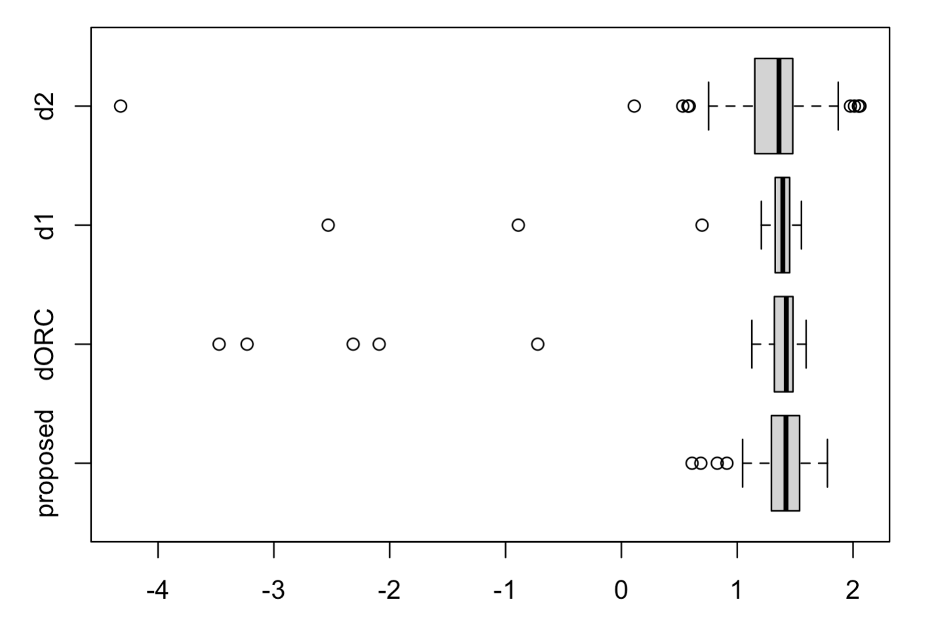}
}
\caption{The original boxplots of values in scenarios L1 with 15 basis functions.}
\label{fig: IFTR l1 15_zoom}
\end{figure}

To assess sensitivity to the number of basis functions when estimating linear IFTR, we conduct additional simulations for scenarios L1 and L2, using 15 cubic B-spline basis functions to expand the coefficient functions. The boxplots of estimated values from four methods for $n=500$ and $n=1,000$ are shown in Figures \ref{fig: IFTR 500_15} and \ref{fig: IFTR 1000_15}. The results obtained using 15 basis functions are similar to those obtained using 7 basis functions. In the L1 scenario, the boxplots for $d1$, $d2$, and $d_{ORC}$ contain several extreme outliers. To facilitate a clearer comparison of the boxplots, we truncate the x-axis limits in these figures. For completeness, the original boxplots with the full x-axis are provided in Figure \ref{fig: IFTR l1 15_zoom}.

\section{Variables used in real data analysis}\label{A: variables details}
The detailed information for variables used in NHANES analysis is listed in table \ref{tab:variables}.

\begin{table}[h] \centering 
\caption{Proxy variables in the NHANES data.} \label{tab:variables} \begin{tabularx}{\textwidth}{p{1.0cm}p{2.4cm}p{5.4cm}X} \toprule Proxy type & Variable & Description & Coding \\ \midrule W & MCQ300a & Close relative had heart attack? & 1 = Yes, 2 = No \\ & MCQ300c & Close relative had diabetes? & 1 = Yes, 2 = No \\ \midrule Z & DEQ034D & Use sunscreen? & 1 = Always, 2 = Most of the time, 3 = Sometimes, 4 = Rarely, 5 = Never \\ \midrule X & BMXBMI & Body mass index & \\ & RIDAGEYR & Age in years & \\ & RIAGENDR & Gender & 1 = Male, 2 = Female \\ & RIDRETH3 & Race & 1 = Mexican American; 2 = Other Hispanic; 3 = Non-Hispanic White; 4 = Non-Hispanic Black; 6 = Non-Hispanic Asian; 7 = Other race \\ & CDQ001 & Ever had pain or discomfort in chest & 1 = Yes, 2 = No \\ & CDQ010 & Shortness of breath on stairs/inclines & 1 = Yes, 2 = No \\ \bottomrule \end{tabularx} \end{table}

\end{document}